\documentclass[aps,prb,showpacs,twocolumn]{revtex4}

\usepackage{graphicx}
\usepackage{amsmath}
\usepackage{amssymb}

\newcommand{\insertpic}[1]{\scalebox{0.34}{\includegraphics{#1}}}
\newcommand{\insertpictwo}[1]{\scalebox{0.34}{\includegraphics[angle=270]{#1}}}

\newcommand{\bvec}[1]{{\mathbf #1}}

\begin{document}

\title{Quantum interference and weak localisation 
effects in the interlayer  magnetoresistance of layered metals}

\author{Malcolm P. Kennett$^{1}$ and Ross H. McKenzie$^{2}$}
\affiliation{$^1$ Physics Department, Simon Fraser University, 8888 University Drive, Burnaby, British Columbia, V5A 1S6, Canada \\
$^2$ Physics Department, University of Queensland, Brisbane 4072, Australia}
\date{\today}

\begin{abstract}
Studies of angle-dependent magnetoresistance oscillations (AMRO) in the
interlayer conductivity of layered metals have generally considered
semi-classical electron transport.  We consider a quantum correction to the
semi-classical conductivity that arises from what can be described as an 
interlayer Cooperon.  This depends on both the disorder potential within
a layer and the correlations of the disorder potential between layers.
We compare our results with existing experimental data on organic 
charge transfer salts that are not explained within the standard semi-classical 
transport picture.  In particular, our results 
may be applicable to effects that have been seen when the applied magnetic 
field is almost parallel to the conducting layers.  We predict the
presence of a peak in the resistivity as the field
direction approaches the plane of the layers.
The peak can occur even when there is
weakly incoherent transport between layers.
\end{abstract}


\maketitle

\section{Introduction}

Angle dependent magnetoresistance oscillations (AMRO) are a valuable 
tool for the investigation of properties of layered metals 
such as organic and cuprate superconductors.
The dependence of the interlayer magnetoresistance on the direction of
the magnetic field has been used to map out a three dimensional Fermi surface
sometimes in exquisite detail
in a diverse range of layered metals. These include
organic charge transfer salts,\cite{kartsovnik-chem}
 strontium ruthenates,\cite{Bergemann,Balicas} 
semiconductor superlattices,\cite{Osada}
monophosphate tungsten bronze,\cite{Beierlein} 
 and an overdoped cuprate.\cite{Hussey,Majed,Majed2} More recent developments 
have extended this to allow one to not only determine the anisotropies 
in the Fermi surface, but also of the anisotropy of scattering on the Fermi surface.
\cite{Majed,Majed2,Analytis,Majed3,Kennett,Kennett2,Singleton}

In addition to the shape of the Fermi surface, the coherence
of interlayer transport in layered metals has been a complicated and 
controversial issue.\cite{millis}
In order for AMRO to be observable, only coherence between neighbouring layers
is required, i.e. weakly incoherent transport, and for almost all angles, this 
leads to AMRO that are equivalent to fully coherent transport
 (i.e. there is a three
dimensional Fermi surface) perpendicular to the planes.\cite{Moses,Kennett}  
In general the two situations can be distinguished in transport measurements 
only if there is a coherence peak in the
resistivity for fields close to parallel to the layers.  In this work, we 
show that quantum interference effects can also lead to a peak in the 
resistivity at fields close to parallel to the layers, for weakly incoherent 
transport.  However, this
peak can be distinguished from the coherence peak by its magnetic field and
temperature dependence, which is determined by the lengthscales and 
timescales over which quantum coherence is destroyed.

The peak in the resistivity for parallel fields that we find arises from a 
contribution to interlayer conductivity in a magnetic field 
$\bvec{B} = B(\sin\theta \cos\phi, \sin\theta \sin\phi, \cos\theta)$
in layered metals which is
a quantum correction to existing semi-classical transport formulae.\cite{Moses,Kennett}
There has been previous work in this 
direction,\cite{Bhatt,Dupuis,Mauz,Yang,gbergmann,szott,raichev} but this 
was either focused on weak
localization in anisotropic metals, or did not consider the magnetic field
configurations relevant for AMRO. 
 Our work is motivated by experiments in which deviations from conventional AMRO are 
seen.\cite{Singleton,SingletonSQUIT,coldea,coldea2,kartsovnik06}  The particular 
deviations of interest are  weak-localization like peaks at
angles near $\theta = 90^\circ$.\cite{SingletonSQUIT}
We calculate the quantum contribution to the interlayer resistivity and show that
it behaves in a similar way in a perpendicular magnetic field to the Cooperon in 
a two dimensional electron system.  We also show that a peak for parallel magnetic
fields emerges naturally for reasonable choices of relevant materials properties,
leading us to suggest that this physics is relevant for a number of recent
experiments.

Our main results are an expression for the interlayer Cooperon correction 
to interlayer conductivity, and asymptotic expressions for magnetic 
fields that are close to parallel to the conducting layers.  We demonstrate the
presence of a small peak in the resistivity for such fields, and discuss the
characteristic behaviour with temperature and perpendicular magnetic field
that allow this feature to be distinguished from coherence peaks at similar
field orientations. 

We note that the calculations we consider here may also be of interest to
those considering weak localization effects in bilayer or multilayer
graphene,\cite{Falko,Savchenko}
semiconductor double quantum wells,\cite{raichev}
semiconductor quantum wells with intersubband scattering,\cite{averkiev}
 and superlattices.\cite{szott}

The paper is structured as follows: in 
Sec.~\ref{sec:Motivation} we briefly review
the physics associated with magnetoresistance
due to weak localisation and discuss the
experiments that motivate our calculations.  In Sec.~\ref{sec:Calc} we give
details of our calculations and results, and in Sec.~\ref{sec:Discussion} 
we discuss the consequences of our results for interpretation of AMRO
experiments, in particular on the issue of coherent versus incoherent interlayer
transport.

\section{Magnetoresistance due to weak localisation}
\label{sec:Motivation}
In order to connect our calculations 
to the experimental data showing unusual peaks in AMRO,
we first review the key physical quantities which determine the
magnitude of weak localization effects in
 two-dimensional metals.\cite{Rammer,Datta}
The total conductivity of the system in a perpendicular
 magnetic field $B_\perp$ can be written as
\begin{equation}
\sigma(B_\perp,T) = \sigma_{cl}(B_\perp,T)+ \delta\sigma(B_\perp,T) ,
\end{equation}
where the first term is the semi-classical contribution and the second term
is associated with quantum interference effects and weak localization.  
The relative magnitude of the two terms is of order
\begin{equation}
\frac{\delta\sigma}{\sigma_{cl}} \sim  \frac{1}{k_F \ell}  ,
\end{equation}
where $k_F$ is the Fermi wave vector and $\ell = v_F\tau$ is the elastic 
mean free path, with $v_F$ the Fermi velocity and $\tau$ the elastic
scattering time. Hence, quantum effects will  be enhanced by
increased disorder, which reduces $k_F \ell$.
At low temperatures in zero magnetic field
\begin{equation}
\delta\sigma(T) \simeq 
-\frac{e^2}{2\pi^2\hbar} \ln\left(\frac{\tau_\phi(T)}
{\tau} \right)  ,
\end{equation}
where $\tau_\phi$ is the inelastic scattering time, which includes all inelastic
processes, e.g. electron-electron and electron-phonon scattering.
Since generally, $ 1/\tau_\phi \sim T^p$, with $p$ some
positive number, the resistivity diverges logarithmically with 
decreasing temperature.

The magnetic field scale (the phase breaking field) at which 
quantum interference effects are destroyed is
$B_\phi=\hbar/(4 e D \tau_\phi)$, where $D=\frac{1}{2} v_F^2\tau$
is the diffusion constant.
The area associated with one quantum of magnetic flux
is related to the Thouless length, $L_{th}\equiv\sqrt{D \tau_\phi}$.
This magnetic field scale can be written in the form
\begin{equation}
 B_\phi(T) = \frac{\hbar}{2e \ell^2}\frac{\tau}{\tau_\phi(T)} ,
\end{equation}
in order to emphasize the fact that this field scale
is temperature dependent.
Destruction of the quantum interference
effects by the magnetic field
leads to a negative magnetoresistance which
is quadratic in the magnetic field for small 
fields.\cite{Hikami,Altshuler,Rammer,Datta}
When 
 $B_0 = \frac{\hbar}{4eD\tau} \gg B_\perp$
and $\tau_\phi \gg \tau$ ($B_0 \gg B_\phi(T)$)
the change in the quantum contribution to the conductivity
due to the magnetic field can be written
in terms of $\psi(z)$,  the digamma function, 
\begin{eqnarray}
\Delta\sigma & \equiv
 & \sigma(B_\perp) -\sigma(B_\perp=0)  \nonumber \\
&= & \frac{e^2}{2\pi^2\hbar} \left\{ \ln\left(\frac{B_\perp}{B_\phi(T)} \right)
 + \psi\left( \frac{1}{2} + \frac{B_\phi(T)}{B_\perp} \right) \right\} , \nonumber \\
& \simeq & 
\frac{e^2}{48\pi^2\hbar} 
\left(\frac{B_\perp}{B_\phi(T)}\right)^2  ,
\label{eq:weakloc_conv}
\end{eqnarray}
where the last equality only holds when  $B_\phi(T)  \gg B_\perp$.
Equation~(\ref{eq:weakloc_conv}) can be used to fit experimental data
for the magnetoconductivity with one
free parameter, $\tau_\phi(T)$, provided the diffusion constant
is known.
For thin metallic films and semiconductor heterostructures
this has proven to be a powerful method for determining the
absolute value and temperature dependence of the 
inelastic scattering rate.\cite{Datta,simmons}

\subsection{Summary of Results}
Our calculations establish the following:          \\
1. When the temperature is low enough that the inelastic
scattering rate is much less than the elastic scattering
rate, i.e. $\tau_\phi \gg \tau$, with no magnetic field
perpendicular to the layers present, there will be a small
increase in the interlayer 
resistivity which is approximately logarithmic 
in temperature. \\
2. At low temperatures there will be a small negative
magnetoresistance on the scale of perpendicular magnetic
fields of order the field $B_a \sim B_\phi$ (defined in
Eq.~(\ref{eqn:Ba})). \\
3. Even when the field parallel to the layers is large there will
   still be a contribution to the resistance due to 
weak localization similar to that which
occurs in zero field. As     
the field is tilted at an angle $\delta \theta$, slightly away  
from the layers, there is a small 
perpendicular field, $B_\perp \simeq B \delta \theta$.
Consequently, a graph of the interlayer
resistance versus $\theta$ will then have a narrow
peak at $\theta =\pi/2$, which has a width proportional to
$B_a/B$ when $\tau_\phi > \tau$.
The temperature dependence of $B_a$ implies the width of the peak
is also  temperature dependent.
The height of the peak relative
to the background magnetoresistance will only depend weakly on
the strength of the field parallel to the layers for fields 
larger than $B_p$ (defined in Eq.~(\ref{eq:bp})).\\
4. The resistance peak is quite distinct from the ``coherence'' peak which
can occur at 
$\theta =\frac{\pi}{2}$ 
if there is coherent interlayer transport (i.e., a three-dimensional 
Fermi surface).\cite{Moses,kartsovnik-chem}
 This coherence peak only exists at very high
fields ($B_\parallel \gg m^* k_F c 
\sqrt{\epsilon_F/t_\perp}/e\tau(T) $), has a width
$\delta \theta = 2 k_F c (t_\perp/ \epsilon_F) $ which is 
independent of temperature and field, and its
height relative to the background scales
with $B$ and $\tau$. \cite{Hanasaki,kartsovnik-chem}\\
5. There are two regimes of parallel field, separated by
an intermediate field scale,
\begin{equation}
B_p \equiv \frac{\hbar}{ec\ell} .
\label{eq:bp}
\end{equation}
For $B_\parallel \ll B_p$, the corrections to the conductivity 
are essentially independent of $B_\parallel$, whereas the 
semiclassical background does depend on $B_\parallel$.
For $B_\parallel  \gtrsim B_p,$ the correction to the conductivity has the same
$B_\parallel$ dependence as the semiclassical contribution, so that
the relative resistivity correction (i.e. the shape of the peak) becomes a 
function of $B_\perp$ only near $\theta = \frac{\pi}{2}.$ 

\subsection{Brief review of experimental results}

Below we discuss several layered materials
in which peaks in the angular dependence of
the interlayer magnetoresistance
have been seen for parallel fields, and discuss whether the
evidence points towards weak localization peaks or coherence peaks.

\subsubsection{$\beta''$-(BEDT-TTF)$_4$AM(C$_2$O$_4$)$_3$Y}

One class of layered metals where weak localization-like peaks have
been observed\cite{coldea2} is the family
$\beta''$-(BEDT-TTF)$_4$AM(C$_2$O$_4$)$_3$Y
of organic charge transfer salts,
 where Y is a solvent molecule, M=Ga,Cr,Fe, and A=H$_3$O$^+$. 
These materials have significant structural
disorder, especially in the anion layers.\cite{coldea}
This can been seen from the fact that the ratio of the resistivity
at room temperature to that at low temperatures is
of order one. In contrast, in many metallic organic
charge transfer salts this 
ratio is as large as one thousand.\cite{Ishiguro}

The Shubnikov-de Haas (SdH) oscillations yield an oscillation with
period 330 T, 
 corresponding to a circular
Fermi surface with wave vector $k_F =1 \, {\rm nm}^{-1}$.
The corresponding Fermi surface area is an
order of magnitude smaller than
in many ET materials.
The small Fermi surface area arises because the
$\beta''$ crystal structure leads to
four bands, three of which are almost completely filled,\cite{canadell}
so the Fermi energy is near the band edge.
The effective mass associated with these SdH oscillations
is close to one free electron mass.
The corresponding Fermi energy is about 35 meV.
The Dingle temperatures, which
are a rough measure of the disorder in the sample,
 estimated from the SdH oscillations
are in the range 1-4 K, up to an order of magnitude
larger than the cleanest organic charge transfer salts.\cite{kartsovnik-chem}

These materials have several features 
unlike other organic charge transfer salts that 
make them more likely to exhibit
weak localization effects:
(a) smaller Fermi wavevector, (b) stronger disorder,
(both (a) and (b) reduce $k_F \ell$),
and (c) the almost 3/4 filling will enhance
the effect of the strong electronic correlations
which narrow the electronic bands, reducing the Fermi energy
$\epsilon_F$.\cite{aside}

In both the M=Ga and Cr (Y=C$_2$H$_2$Cl$_2$) materials in zero field
the interlayer resistivity increases approximately logarithmically
with temperature below about 20 K.\cite{coldea2}
The total change is about one per cent in the temperature range 1-20 K.
At a temperature of 1.5 K, perpendicular fields of about 0.2 T and 2 T,
destroy this feature in the M=Ga and M=Cr materials respectively.
Furthermore, as the field is tilted towards
the plane of the layers there is a small peak
at $\theta =\pi/2$
 in the
interlayer resistance versus angle
for both the M=Cr and M=Ga material.
 The width of this peak,  
for the M=Cr material,
$\delta\theta$ (in radians), has the field dependence              
$ B_x/B$, with $B_x=0.25$ T, between about
3 and 14 T.\cite{coldea2}
Note that this is the same scale on which the weak localisation
peak occurs for $\theta=0$ and small fields.
The magnitude of the peak is about 0.3 per cent
of the background magnetoresistance and
depends weakly on the parallel field.
As the temperature increases from
0.7 K to 3 K the peak becomes broader and smaller,
and is not visible at 6 K, whereas the background
magnetoresistance changes little in the same temperature range.

\subsubsection{$\alpha$-[(ET)$_{1-x}$(BETS)$_x$]$_2$KHg(SCN)$_4$}

In $\alpha$-[(ET)$_{1-x}$(BETS)$_x$]$_2$KHg(SCN)$_4$ with $x \simeq 0.03$,
the resistance versus temperature  curve shows an upturn below  about 5 K.
\cite{kartsovnik} About ten per cent of this growth can be suppressed with a magnetic field.
The peak that occurs at $\theta =\pi/2$ is                            
about 50  times broader than the coherence peak that occurs in clean 
$\alpha$-(ET)$_2$KHg(SCN)$_4$ samples at fields of order a few tesla.\cite{kartsovnik06}
For low fields, up to 1.5 T, the shape of the peak depends only on $B_\perp$, 
as we would expect for weak localization behaviour, and at higher fields, the 
peak becomes broader than at low fields.  The high field behaviour that is not 
captured in our theory may be related to approximations we use that should break
down at very large fields (e.g., 15 T in this case).
In contrast, the
coherence peak for $\alpha$-(ET)$_2$KHg(SCN)$_4$
has an absolute height that increases monotonically
with increasing magnetic field.\cite{kartsovnik06,kartsovnik}

\subsubsection{$\kappa$-(BEDT-TTF)$_2$Cu(NCS)$_2$}

The magnitude and temperature dependence of the inelastic
scattering rate
was recently determined in this material using
AMRO.\cite{Singleton}
The data was fit to the form
\begin{equation}
\frac{1}{\tau(T)}
=\frac{1}{\tau_{11}} + AT^2 ,
\label{eq:tau_temp}
\end{equation}
with $\tau_{11} \simeq $ 3 ps and $ A \simeq $ 0.006 ps$^{-1}$K$^{-2}$.
This temperature dependence is also consistent with that of
the dc resistivity in many organic charge transfer salts.\cite{Powell}
The first term is associated with elastic scattering due to
disorder.
The quadratic temperature dependence can be associated with
inelastic scattering due to
electron-electron interactions in a Fermi liquid.
The ratio of the inelastic scattering time $\tau_\phi(T)$ to
the intralayer elastic scattering time is
$\tau_\phi/\tau_{11}=50 ({\rm K}/T)^2$.
Thus, as the temperature increases
from 1 K to 20 K the ratio decreases from 50 to 0.1
and 
$\tau_\phi/\tau_{11}$ will decrease from about 1 to 0.1.

We also suggest that weak localisation may be the origin of the feature near
$\theta =\frac{\pi}{2}$ that Singleton {\it et al}.\cite{Singleton} assigned to the coherence
peak.  The peak width appears to broaden with increasing temperature 
(as would be expected for decreasing $\tau_\phi$, and hence increasing $B_\phi(T)$).
In earlier data on the same material,\cite{SingletonSQUIT}
the peak height drops with increasing temperature and the peak width increases
with increasing temperature.  For an inelastic scattering rate that 
increases with temperature
this is to be expected for a weak localisation
peak, since $B_\phi(T)$ increases with increasing temperature
and the width of the peak goes as $B_\phi$, 
whilst its height should increase logarithmically with temperature.

\subsubsection{Intercalated graphite}
A peak near parallel field has also been seen
in intercalated graphite materials which display
unusual inverted AMRO.\cite{enomoto} However, the width of the peak appears to 
be independent of magnetic field which would tend to argue against weak
localization effects.

\section{Model and Calculations}
\label{sec:Calc}

\subsection{Interlayer charge transport}

In previous work, \cite{Kennett} we considered the AMRO that arise when 
there is an anisotropic Fermi surface and anisotropic scattering, 
in two different limits.  These two limits were the limit of coherent 
interlayer transport and the limit of weakly incoherent inter-layer 
transport in which there is hopping between adjacent layers.  We shall not
consider anisotropy in the Fermi surface, scattering or the interlayer
hopping here, as these are additional complications beyond the physics 
that is our main interest.  For isotropic 
weakly incoherent interlayer transport the
 Hamiltonian describing interlayer charge transport is 
\begin{equation}
{\mathcal H} = t_\perp \sum_{i,j=1,2} \left[c_i^\dagger c_j + h.c.\right], 
\end{equation}
where $t_\perp$ is the interlayer hopping integral.
We assume that we are in weakly incoherent regime such that
\begin{equation}
t_\perp \ll {1 \over \tau} .
\end{equation}
There is no observable difference in the AMRO between the
coherent and weakly incoherent models of interlayer transport 
 except for polar angles of the magnetic field 
very close to 90$^\circ$, corresponding to a magnetic field parallel to 
the planes.\cite{Kennett,Moses}
The interlayer conductivity that is deduced from such a hopping
Hamiltonian is:\cite{Moses}

\begin{eqnarray}
\sigma_\perp & = & \frac{e^2 c t_\perp^2}{\hbar \pi L^2} \int d^2\bvec{r}_1 
\int d^2\bvec{r}_2 \left< G_1^R(\bvec{r}_1,\bvec{r}_2) G^A_2(\bvec{r}_2,
\bvec{r}_1) \right. \nonumber \\
& & \left. \hspace*{2.0cm} + G_1^A(\bvec{r}_1,\bvec{r}_2) G^R_2(\bvec{r}_2,
\bvec{r}_1) \right> ,
\end{eqnarray}
where $c$ is the interlayer spacing, $L$ is the sample size, and 
$G_i^{R(A)}$ is the retarded (advanced) Green's function in layer $i$.
The angle brackets $\left< \ldots \right>$ indicate an average over disorder.
Generically in calculating the disorder-averaged conductivity, there 
are two classes of diagrams to consider, the ladder diagrams that lead to the 
``Diffuson'', and the maximally crossed diagrams that lead to the 
``Cooperon''.\cite{Datta,Rammer}
In our previous work, we found that the
AMRO from weakly incoherent interlayer transport could be derived using a 
``Diffuson''-like equation.\cite{Kennett}  We expect that there should
also be a contribution to the inter-layer conductivity 
from a ``Cooperon'' like process, that will have a
different magnetic field dependence to the ``Diffuson'' term due
to their differing time-reversal properties.  
It is this question that we address in this paper, and we find that
indeed, an extra quantum correction to the interlayer conductivity is generated.
Moreover, it has distinct experimental signatures.

\subsection{Impurity correlations in a layered metal}

We now discuss the properties of the probability distribution for
impurities in a layered metal, and their consequences for impurity correlations.
We assume that the probability distribution $p(\bvec{r})$ for the locations of
impurities can be factorized into separate distributions describing the
distributions parallel ($p_\parallel(x,y)$) and perpendicular ($p_z(z)$) to
the layers\cite{szott}
$$p(\bvec{r}) = p_\parallel(x,y)p_z(z).$$
We assume that  an individual impurity (or lattice defect)
located at $\bvec{R}$ has a potential
$V_{\rm imp}(\bvec{r} - \bvec{R})$ associated with it, and that
$$\left<V(\bvec{r})\right> = n_i\int d^3\bvec{r}^\prime p(\bvec{r}^\prime)
V_{\rm imp}(\bvec{r} - \bvec{r}^\prime) = 0.$$
Correlations in the random potential take the form
\begin{eqnarray}
\left<V(\bvec{r})V(\bvec{r}^\prime)\right>
& = &  n_i\int d^3\bvec{r}^{\prime\prime} p(\bvec{r}^{\prime\prime}) 
V_{\rm imp}(\bvec{r} - \bvec{r}^{\prime\prime}) 
V_{\rm imp}(\bvec{r}^\prime - \bvec{r}^{\prime\prime}) \nonumber \\
& = & n_i \int d^2\bvec{r}^{\prime\prime}
p_\parallel(x^{\prime\prime},y^{\prime\prime}) \int dz^{\prime\prime}
p_z(z^{\prime\prime}) \nonumber \\
& & \hspace*{0.5cm} \times V_{\rm imp}(\bvec{r} - \bvec{r}^{\prime\prime})
V_{\rm imp}(\bvec{r}^\prime - \bvec{r}^{\prime\prime}). \nonumber \\
& & 
\end{eqnarray}
We shall assume that there is a uniform distribution of impurities in the
planes parallel to the conducting layers, i.e. $p_\parallel(x,y) = \frac{1}{L^2}$,
where $L$ is the sample size.  All structure in the impurity distribution is 
thus in $p_z$ which will tend to be peaked in the regions in between conducting
layers.  We note that 
periodicity and inversion symmetry imply that $p(z) = p(-z) = p(z+c)$.

We give a layer index to the random potential $V_j$ depending on whether
the electron $z$ co-ordinate lies in layer 1 or layer 2.  There will then be
two important types of disorder correlations $\left< V_i(\bvec{r}^\prime) 
V_j(\bvec{r}) \right>$ where $i$ and $j$ are layer indicies. 
Firstly, correlations within a single
layer, $\left< V_1(\bvec{r}^\prime) V_1(\bvec{r}) \right>$, which contribute to the
in-plane elastic scattering rate, and appear in the disorder-averaged single 
layer Green's functions. Secondly, correlations between layers
$\left< V_1(\bvec{r}^\prime) V_2(\bvec{r}) \right>$, that are relevant for
interlayer conductivity.  The corresponding expressions are (using the symmetry
properties of $p_z$):

\begin{eqnarray}
\left< V_1(\bvec{r}^\prime) V_1(\bvec{r}) \right>
& = & \frac{n_i}{L^2} \int d^2\bvec{r}^{\prime\prime}
\int_{-{\frac{c}{2}}}^{\frac{c}{2}} dz \, p_z(z) \nonumber \\
& & \hspace*{0.3cm} \times V_{\rm imp}(\bvec{r}^\prime - \bvec{r}^{\prime\prime}, z)
V_{\rm imp}(\bvec{r} - \bvec{r}^{\prime\prime}, z) ,
\label{eq:voneone} \\
\left< V_1(\bvec{r}^\prime) V_2(\bvec{r}) \right>
& = & \frac{n_i}{L^2} \int d^2\bvec{r}^{\prime\prime} \int_0^c dz \, p_z(z)
\nonumber \\
& & \hspace*{0.1cm} \times
V_{\rm imp}(\bvec{r}^\prime - \bvec{r}^{\prime\prime}, c-z) 
 V_{\rm imp}(\bvec{r} - \bvec{r}^{\prime\prime}, -z) , \nonumber \\
\end{eqnarray}

\subsection{Elastic scattering times in a layered metal}

In order to be clearer 
as to what we mean by an interlayer Cooperon,
we now discuss the nature of the disorder potential
and the associated elastic scattering times.
We shall assume that disorder correlations 
can be represented as delta correlated, i.e.,
\begin{eqnarray}
\left< V_i(\bvec{r}^\prime) V_j(\bvec{r}) \right> = \bar{U}_{ij}^2
\delta(\bvec{r} - \bvec{r}^\prime),
\end{eqnarray}
where $i,j = 1,2$. It is useful to rewrite
\begin{eqnarray}
\bar{U}_{ij}^2 = \frac{\hbar^3}{m^*\tau_{ij}} =
\frac{\hbar}{2\tau_{ij}} \frac{1}{\pi N_s},
\end{eqnarray}
 where $N_s = \frac{m^*}{2\pi \hbar^2}$ is the two-dimensional
 density of states, not including spin degeneracy,
 and $m^*$ is the effective mass. 
Thus there are two distinct elastic scattering times associated with the disorder.
$\tau_{11} = \tau_{22}$ is the scattering time for in-plane elastic
scattering, whilst $\tau_{12} = \tau_{21}$ is the scattering time corresponding to scattering correlations between adjacent layers. 
If the impurities are all equidistant between the layers, i.e., $p(z)$ has
a maximum at $z = \pm c/2$, then we will have $\tau_{11} \simeq
\tau_{12}$ , whereas if the impurities are located in the conducting layers, 
we expect $\tau_{11} < \tau_{12}$.

We note that previous theoretical treatments of AMRO
have not considered the possible role of disorder 
correlations between neighbouring layers which
could also modify the semi-classical interlayer
conductivity.\cite{smith}

\begin{figure}[htb]
\insertpic{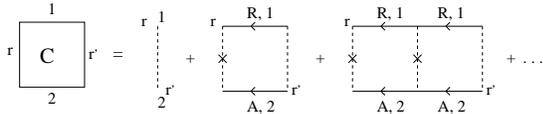}
\caption{Diagrams that give rise to the ``interlayer Cooperon'',
 described by Eq.~(\ref{eq:int-coop}).
Neighbouring layers are indicated as 1 and 2. A and R denote advanced
and retarded Greens functions, respectively.}
\label{fig:coop}
\end{figure}

\subsection{Interlayer Cooperon in the absence of magnetic field}               

\begin{figure}[htb]
\insertpic{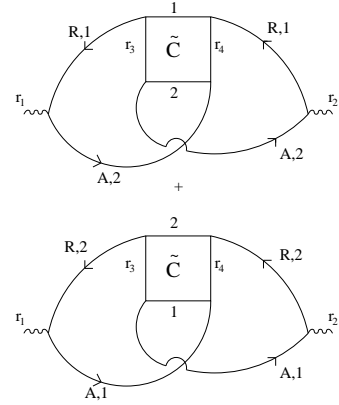}
\caption{Diagrams for the  quantum correction
to the interlayer conductivity.  Note that $\tilde{C} = \bar{U}_{12}^2 C$
is related to the interlayer Cooperon.
This is a diagrammatic representation of Eq.~(\ref{eq:wl}).
}
\label{fig:cond_diag}
\end{figure}
The conductivity diagrams corresponding
to the situation we consider are as shown in Fig.~\ref{fig:cond_diag}.
Following the treatment of the two-dimensional problem by
Rammer,\cite{Rammer} or Datta,\cite{Datta} we can write the
correction to the interlayer
conductivity from the Cooperon (in zero magnetic field) as 
\begin{eqnarray}
\delta\sigma^C_\perp & = &  
\frac{e^2 c t_\perp^2}{\pi\hbar L^2}
\int d\epsilon \left(-\frac{\partial f}{\partial\epsilon}\right)
\int d^2\bvec{r}_1 \int d^2\bvec{r}_2
\int d^2\bvec{r}_3 \int d^2\bvec{r}_4 \nonumber \\
& & \times \left\{
G_1^R(\bvec{r}_1,\bvec{r}_3) G_1^R(\bvec{r}_4,\bvec{r}_2) \bar{U}_{12}^2 
C_{12}(\bvec{r}_3,\bvec{r}_4) \right. \nonumber \\
& & \times \left. G_2^A(\bvec{r}_2,\bvec{r}_3)
G_2^A(\bvec{r}_4,\bvec{r}_1)  + 
G_1^A(\bvec{r}_4,\bvec{r}_1) G_1^A(\bvec{r}_2,\bvec{r}_3) \right. \nonumber 
\\ & & \left. \times \bar{U}_{12}^2
C_{21}(\bvec{r}_3,\bvec{r}_4) G_2^R(\bvec{r}_1,\bvec{r}_3)
G_2^R(\bvec{r}_4,\bvec{r}_2)\right\} ,
\label{eq:wl}
\end{eqnarray}
where $\tilde{C} = \bar{U}^2_{12} C$ 
is the Cooperon [Fig.~\ref{fig:coop}].  The 
diagrammatic expansion of the Cooperon in Fig.~\ref{fig:coop} implies
the following equation for the interlayer Cooperon:

\begin{eqnarray}
C_{12}(\bvec{r},\bvec{r}^\prime) = \delta(\bvec{r} - \bvec{r}^\prime)
 + \int d^2\bvec{r}^{\prime\prime} 
\tilde{J}_{12}^C(\bvec{r},\bvec{r}^{\prime\prime})
C_{12}(\bvec{r}^{\prime\prime},\bvec{r}^\prime),
\label{eq:int-coop}
\end{eqnarray}
where 
\begin{eqnarray}
\tilde{J}_{12}^C(\bvec{r},\bvec{r}^{\prime}) = \bar{U}_{12}^2 
G^R_1(\bvec{r},\bvec{r}^\prime) G^A_2(\bvec{r},\bvec{r}^\prime) ,
\end{eqnarray}
is the Cooperon insertion.  We can Taylor expand 
$C_{12}(\bvec{r}^{\prime\prime},\bvec{r})$ and obtain \cite{Rammer}
the following generalization of the standard equation for the Cooperon
in the experimentally relevant limit that 
the in-plane elastic scattering time, $\tau_{11}$, is much less than the
inelastic scattering time, $\tau_{11} \ll \tau_{\phi}$:
\begin{eqnarray}
\left\{\frac{1}{\tau_a} - D_{a}\nabla^2_r \right\}  
C_{12}(\bvec{r},\bvec{r}^\prime) = \frac{1}{\tau_{11}} \delta(\bvec{r} -
\bvec{r}^\prime),
\label{eq:cooperon_equation_one}
\end{eqnarray}
which has the same form as the equation for the intralayer
Cooperon, except the diffusion constant is
\begin{eqnarray}
D_{a} = \frac{\tau_{11}}{\tau_{12}} D_0 =       
 \frac{v_F^2 \tau_{11}^2}{2 \tau_{12}}
\end{eqnarray}
and the scattering rate which destroys the 
interlayer Cooperon is
\begin{eqnarray}
\frac{1}{\tau_a(T)} = 
\frac{1}{\tau_{11}} 
- \frac{1}{\tau_{12}} 
+ \frac{1}{\tau_{\phi}(T)} 
\label{eq:tau_a}
\end{eqnarray}
In contrast, the first two terms do not appear for the intralayer Cooperon.
For a general impurity potential $\tau_{11} < \tau_{12}$ 
and so the effective dephasing rate will have a non-zero value at zero temperature.
The structure of this
expression has some similarities
 to what occurs with spin-orbit scattering\cite{Hikami} or
intervalley scattering.\cite{Falko}

\subsection{Interlayer Cooperon in the presence of a magnetic field}

In the presence of a magnetic field parallel
to the layers we note that the Green's functions
in layers 1 and 2 are related by a gauge factor\cite{Moses}
$$G_1(\bvec{r},\bvec{r^\prime}) = 
\exp\left(\frac{iec}{2\hbar}
\bvec{B_{\parallel}}\cdot(\bvec{r} - \bvec{r}^\prime) \right)
G_2(\bvec{r},\bvec{r^\prime}),$$
and that the Green's functions can be written as a gauge dependent phase
multiplying a gauge invariant piece:
$$G(\bvec{r},\bvec{r}^\prime) =
\exp
\left(\frac{ie}{\hbar}\int_{\bvec{r^\prime}}^\bvec{r}\bvec{A}_\perp\cdot d\bvec{l}
\right)G_0(\bvec{r},\bvec{r}^\prime),$$
where $\bvec{A}_\perp$ is the vector potential associated
with the perpendicular magnetic field
 $\bvec{B_{\perp}}$ and then
\begin{eqnarray}
\tilde{J}_{12}^C(\bvec{r},\bvec{r}^\prime) 
&\to& \tilde{J}_{12}^{C,0}(\bvec{r},\bvec{r}^\prime) 
\exp\left(\frac{iec}{2\hbar}\bvec{B_{\parallel}}
\cdot(\bvec{r} - \bvec{r}^\prime)\right) \nonumber \\
& & \hspace*{0.5cm} \times
\exp\left(\frac{2ie}{\hbar} (\bvec{r} 
- \bvec{r}^\prime)\cdot\bvec{A}_\perp(\bvec{r})\right) .
\end{eqnarray}
This allows us to generalize Eq.~(\ref{eq:cooperon_equation_one}), as follows.
We choose the Landau gauge to 
determine the perpendicular field, 
make a gauge transformation to eliminate the $\bvec{B}_\parallel$
dependent term, and end up with an equation for 
$C_{12}^\prime(\bvec{r},\bvec{r}^\prime)
 = \exp\left(\frac{iec}{2\hbar}\bvec{B_{\parallel}}\cdot(\bvec{r} - \bvec{r}^{\prime})\right)
C_{12}(\bvec{r},\bvec{r^\prime})$ which is (when we approximate $\tilde{J}^{C,0}_{12}$
by its zero field value)

\begin{eqnarray}
\left\{\frac{1}{\tau_{a}} - D_{a}\left(\nabla_r - \frac{2ie}{\hbar}\bvec{A}_\perp
\right)^2 \right\} C_{12}^\prime(\bvec{r}^\prime,\bvec{r}) = 
\frac{1}{\tau_{11}} \delta(\bvec{r} - \bvec{r}^\prime) . \nonumber \\
\label{eq:cooperon_equation_two}
\end{eqnarray}
Note that the approximation of $\tilde{J}^{C,0}_{12}$ by its zero field
value implies that for high enough fields, our expressions will not be applicable.
Now, $C_{12}(\bvec{r}_3, \bvec{r}_4) \simeq C_{12}(\bvec{r}_3, \bvec{r}_3)$,
since the Cooperon is dominated by loops of time-reversed paths,
and in a perpendicular magnetic field $B_\perp$, the solution of equations of the 
form of Eq.~(\ref{eq:cooperon_equation_two}) is well known.\cite{Altshuler,Rammer}
For our parameters this is:
\begin{eqnarray}
C_{12}^\prime(\bvec{r}_3,\bvec{r}_3)
= \frac{2eB_\perp}{2\pi\hbar} \sum_{n=0}^{n_{\rm max}}
\frac{1}{4D_{a}|e|B_\perp\frac{\tau_{11}}{\hbar}(n+\frac{1}{2}) +
\frac{\tau_{11}}{\tau_{a}}}, \nonumber \\
\label{eq:coop_perp}
\end{eqnarray}
with $n_{\rm max} \simeq l_B^2/\ell^2$,\cite{Rammer} where $\ell = v_F\tau_{11}$
is the in-plane mean free path.

To calculate the conductivity, we need
to take into account the phase acquired by the propagators 
due to the vector potential differing in each layer.  Thus,
considering the first term in the conductivity (since the second term
is its complex conjugate) and switching to momentum space propagators the expression
simplifies to:
\begin{eqnarray}
\delta\sigma^C_\perp & = &  2 {\rm Re}\left[
\frac{e^2 c t_\perp^2 \bar{U}_{12}^2 }{\pi\hbar L^2}
\int \frac{d^2\bvec{k}_1}{(2\pi)^2} \frac{d^2\bvec{k}_2}{(2\pi)^2}
G^R(\bvec{k}_1) G^R(\bvec{k}_2) \right. \nonumber \\
& & \left. \times \,
G^A\left(\bvec{k}_1 + \frac{ec}{2\hbar}\bvec{B}_\parallel\right)
G^A\left(\bvec{k}_2 + \frac{ec}{2\hbar}\bvec{B}_\parallel\right) \right.
\nonumber \\ & & \left. \times
\int d\bvec{r}_3 \int d\bvec{r}_4 C_{12}(\bvec{r}_3,\bvec{r}_4)
e^{i(\bvec{k}_1 + \bvec{k}_2)\cdot
(\bvec{r}_3 - \bvec{r}_4)} \right] . \nonumber \\
\end{eqnarray}
Making use of Eq.~(\ref{eq:coop_perp}) we obtain the expression:
\begin{eqnarray}
\delta\sigma^C_\perp =  \frac{2ce^2 t_\perp^2 \bar{U}_{12}^2}{\pi\hbar}
\left(\frac{eB_\perp}{\pi\hbar}\right) \sum_{n=0}^{n_{\rm max}}
\frac{{\rm Re} [F(\bvec{B}_\parallel)]}{4D_a|e|
B_\perp\frac{\tau_{11}}{\hbar}(n+\frac{1}{2})
+ \frac{\tau_{11}}{\tau_a}}. \nonumber \\
\label{eq:Csum}
\end{eqnarray}
In Appendix~\ref{sec:Fbparallel} we show that for 
$ ecB_\parallel \ll     \hbar k_F $ 
(the experimentally relevant limit)
\begin{eqnarray}
{\rm Re}[F(\bvec{B}_\parallel)] & \simeq &
 -\frac{2m^*
\tau_{11}^3}{\hbar^5}f\left(\frac{e}{\hbar}B_\parallel c\ell\right) ,
\label{eq:fbparallel_two}
\end{eqnarray}
where 
\begin{eqnarray}
f(x) \equiv  \frac{1 + \frac{x^2}{8}}{(1+\frac{x^2}{4})^{3/2}}.
\label{eq:fx}
\end{eqnarray}
Note that $f(x) \simeq 1 - \frac{x^2}{4}$ for $x \ll 1$ 
and $f(x) \simeq \frac{1}{x}$ for $x \gg 1$.  

We can rewrite Eq.~(\ref{eq:Csum}) 
in a similar form to the magnetoresistance in a
thin film\cite{Altshuler} using the properties of the digamma function
$\psi(x)$,\cite{GnR} which is related to the sum we have obtained via
$$\psi(x+n+1) - \psi(x) = \sum_{m=0}^{n} \frac{1}{x+m}.$$ 
Hence we obtain
\begin{eqnarray}
\delta\sigma^C_\perp(B_\perp) & = & 
-\sigma_1 \left[\psi\left(\frac{B_a}{B_\perp} + n_{\rm max}
 + \frac{3}{2} \right) - 
\psi\left(\frac{B_a}{B_\perp} + \frac{1}{2}\right)\right],
 \nonumber \\
& \simeq & \sigma_1 \left[\ln\left(\frac{B_\perp}{B_0}\right) + 
\psi\left(\frac{1}{2} + \frac{B_a}{B_\perp}\right)\right], 
\label{eq:sigma_correction}
\end{eqnarray}
with $B_0 = \frac{\hbar}{e\ell^2}$, and
\begin{eqnarray}
 \sigma_1(B_\parallel) = \frac{2ce^2 t_\perp^2}{\pi^2\hbar^3 v_F^2} 
f\left(\frac{B_\parallel}{B_p}\right) ,
\end{eqnarray}
where $B_p$ is the field scale defined in Eq.~(\ref{eq:bp}).

The magnetic field dependence is of the same form as that for
the weak localization correction to intralayer conductivity, Eq.~(\ref{eq:weakloc_conv}),
with $B_\phi(T)$ replaced by the magnetic field scale
\begin{eqnarray}
B_a(T) \equiv \frac{\hbar}{4eD_a\tau_a(T)},
\label{eqn:Ba}
\end{eqnarray}
where $\tau_a$ is given by Eq.~(\ref{eq:tau_a}).
Consequently, we note that unlike the phase breaking field
associated with the intralayer conductivity
that this quantity will be non-zero even at zero temperature.
In the absence of a perpendicular magnetic field
the interlayer Cooperon correction to conductivity reduces to
\begin{eqnarray}
\delta\sigma^C_\perp (B_\perp = 0, T)  =  
 \sigma_1 
\ln\left(\frac{\tau_{12}}{\tau_a(T)}\right) .
\end{eqnarray}
Thus, the interlayer resistivity increases logarithmically
as the temperature decreases.

We now compare these quantum
corrections to
the  semi-classical
conductivity which describes
conventional AMRO.
The latter is given by\cite{Moses}
\begin{equation}
\sigma_{\rm conv}(\bvec{B}) = \sigma_0 \sum_{n=-\infty}^\infty 
\frac{J_n(\gamma\tan\theta)^2}{1 + (n\omega_0\tau_{11}\cos\theta)^2} ,
\label{eq:sigmaconv}
\end{equation}
where $\gamma = ck_F$, $J_n(z)$ is the Bessel function of order $n$, 
 $\omega_0 = \frac{eB}{m^*}$, and the conductivity
at zero field and at $\theta=0$ is
\begin{equation}
\sigma_0 = \frac{2ce^2 t_\perp^2 m^* \tau_{11}}{\pi\hbar^4}.
\label{eq:sigma0}
\end{equation}
 We can estimate the relative size of the
quantum correction to the conventional term as
\begin{equation}
\frac{\sigma_1}{\sigma_0} =  
\frac{f\left(\frac{e}{\hbar}B_\parallel c\ell\right)}{\pi k_F \ell} 
\sim \frac{1}{k_F \ell},
\end{equation}
for fields such that $eB_\parallel c \ell/\hbar \ll 1$. 
 This indicates that the
effects we discuss here are most likely to be seen in ``dirty''   
samples (i.e., short mean-free path) 
and with a small Fermi surface.

\subsection{Peak in the vicinity of $\theta = \pi/2$}
Now, $n_{\rm max}$ depends on polar angle -- as $\theta \to \frac{\pi}{2}$,
$B_\perp \to 0$ and hence $n_{\rm max} \to  \infty$.
Using the 
asymptotic expansion of $\psi(x)$ at large $x$, we obtain (in the limit that 
$\frac{B_0}{B_a} \sim \frac{\tau_{a}}{\tau_{11}} \gg 1$)

\begin{eqnarray}
\delta \sigma^C & \simeq & -\sigma_1 
\left\{ \ln\left(\frac{B_0}{B_a}\right) - \frac{B_\perp^2}{24B_a^2} \right\} .
\end{eqnarray}
This correction will have two different dependences on parallel field 
arising from $f\left(\frac{e}{\hbar}B_\parallel c\ell\right)$.  At fields
much less than $B_p = \hbar /ecl$ it will be independent of parallel 
field, whilst for fields larger than $B_p$ it takes the form

\begin{eqnarray}
\delta \sigma^C & \simeq & -
\frac{\sigma_0}{\pi k_F \ell   }
\frac{B_p}{B_\parallel}
\left\{ \ln\left(\frac{B_0}{B_a}\right) - \frac{B_\perp^2}{24B_a^2} \right\} .
\end{eqnarray}

In the limit that $\theta \to \frac{\pi}{2}$, we can also compare with the limit from $\sigma_{\rm conv}$ which can be obtained by standard saddle point methods applied to
an integral representation of Eq.~(\ref{eq:sigmaconv}).
In the limit that $\gamma\tan\theta \gg 1$,\cite{Moses}
$$ \sigma_{\rm conv} \simeq \sigma_0 \frac{B_p}{B_\parallel}
\left[ 1 + 2\exp\left(-\frac{\pi}{\omega_0\tau_{11}\cos\theta}\right)
\sin(2\gamma\tan\theta) \right].$$
If we compare the two contributions in the limit that
 $\theta \to \frac{\pi}{2}$,
for fields larger than $B_p$, then we find that 
the size of the quantum correction 
relative to the semiclassical magnetoconductivity
is the same as for the zero field correction and has the
same temperature dependence
\begin{equation}
\frac{\delta\sigma^C}{\sigma_{\rm conv}} \simeq - 
\frac{1}{\pi k_F \ell}
\ln\left(\frac{\tau_{12}}{\tau_a(T)}\right) .
\end{equation}

In order to compare with experiment, 
it is useful to separate the conductivity as
$\sigma(\bvec{B}) = \sigma_{\rm back}(\bvec{B}) + \Delta\sigma(\bvec{B}),$
where $\sigma_{\rm back}(\bvec{B}) = \sigma_{\rm conv}(\bvec{B}) + 
\delta\sigma^C(B_\perp = 0),$ since it is only the negative magnetoresistance
for increasing $B_\perp$ that will be visible.  Defining 
$\rho_{\rm back} = 1/\sigma_{\rm back}$, we have
$$ \frac{\rho_{zz}}{\rho_{\rm back}} 
\simeq 1 - \frac{\Delta\sigma^C}{\sigma_{\rm back }},$$
which implies a decrease in the resistivity,
which grows as $B_\perp$ increases, i.e. as $\theta$ deviates from  $\frac{\pi}{2}$.
In consequence, there will be a peak above the background resistance at 
$\theta = \frac{\pi}{2}$.

\section{Discussion}
\label{sec:Discussion}

The correction to the resistivity at $\theta = \frac{\pi}{2}$ we found in the previous section 
is much smaller in magnitude than the conventional term, but since conventional AMRO
are featureless in this region, this quantum contribution can still be
visible above the background. The Cooperon should lead to a small
peak in the resistivity in the vicinity of $\theta = \frac{\pi}{2}$.
The values of $\theta$ for which this should be 
true are those such that both asymptotic formulae hold, i.e., $\cos\theta \ll 
1/(ck_F)$ and $B\cos\theta = B_\perp \ll B_a$.

\subsection{Numerical Results}

\begin{figure}[htb]
\insertpictwo{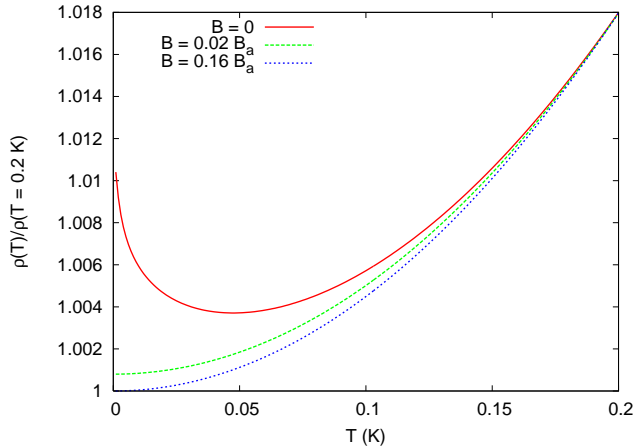}
\caption{Interlayer resisitivity as a function of 
temperature for several different magnetic fields perpendicular
to the layers. In the absence of the field the
resistivity increases logarithmically with temperature
due to weak localization effects.
This feature disappears as the magnetic field
becomes comparable to the field $B_a$.
The resistivity is normalized to its value
at 0.2 K.
The parameters used in these plots are $k_F = 1.4 \, {\rm nm}^{-1}$, 
$c = 2 \, {\rm nm}$, $\tau_{11} = 2$~ps, $\tau_{12} = 1.01 \tau_{11}$,
$\epsilon_F  = 40$ meV, and $\tau_\phi(T)/\tau_{11} = 50(K/T)^2$.
For these parameter values $B_a(T=0)= 0.002$ T.}
\label{fig:temp}
\end{figure}

\begin{figure}[htb]
\insertpictwo{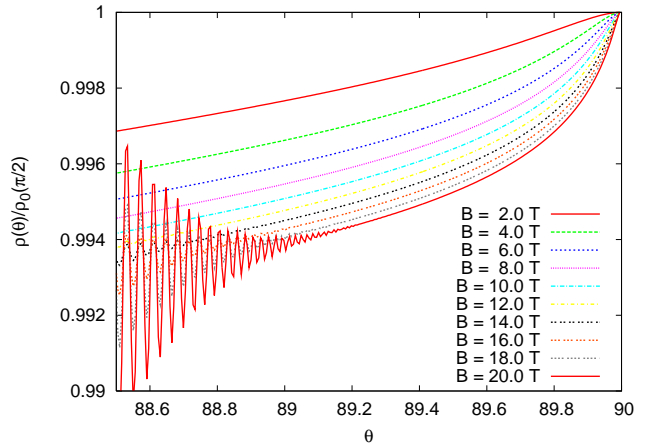}
\caption{Interlayer resisitivity as a function of 
the angle $\theta$, between the magnetic field and the normal to
the layers. The peak at 90 degrees is due to weak localisation
effects. Note that it is only about 0.5 percent of the total resistivity. 
The resistivity is normalized to the sum of the semi-classical resistivity
and the zero field quantum correction. 
As the magnetic field increases 
from $2$ to $20$ T, the size of the peak decreases and its width increases.
The parameters used in these plots are $k_F = 1.4 \, {\rm nm}^{-1}$, 
$c = 2 \, {\rm nm}$, $\tau_{11}  = 2$ ps, $\tau_{12} = 1.01 \tau_{11}$
$\epsilon_F  = 40$ meV, and $\tau_a = 50$ ps.}
\label{fig:peak}
\end{figure}

We now evaluate the resistivity correction obtained
in the previous section, Eq.~(\ref{eq:sigma_correction}), 
for  parameters appropriate for  $\alpha$-(ET)$_2$KHg(SCN)$_4$
 (Refs.~\onlinecite{kartsovnik-chem,kartsovnik06}).
Fig.~\ref{fig:temp} shows the calculated
temperature dependence and  Fig.~\ref{fig:peak}
the angular dependence.
We calculate the dependence of the
interlayer resistivity on temperature, field, and field direction.
We choose $k_F = 1.4 \, {\rm nm}^{-1}$, $\epsilon_F  = 40$~meV, 
$\tau_{11} = 2$ ps, $\tau_{12} = 1.01 \tau_{11}$  and $c \simeq 20$ \AA. 
We choose $\tau_\phi = 66$ ps, which gives $\tau_a = 50$ ps, a reasonable 
value based on the discussion following Eq.~(\ref{eq:tau_temp}).
From Eq.~(\ref{eq:tau_temp}) we see that $\tau_\phi \sim $ 66 ps at $T \simeq 1.2$ K.
These choices correspond to fields of $B_0 \sim 0.2$ T and $B_a \sim 0.002$ T.
 The width of the
peak is approximately $\Delta\theta = 0.5^\circ$ 
and the height of the peak is about $0.5\%$ of the background resistance at all
fields.  

\begin{figure}[htb]
\insertpictwo{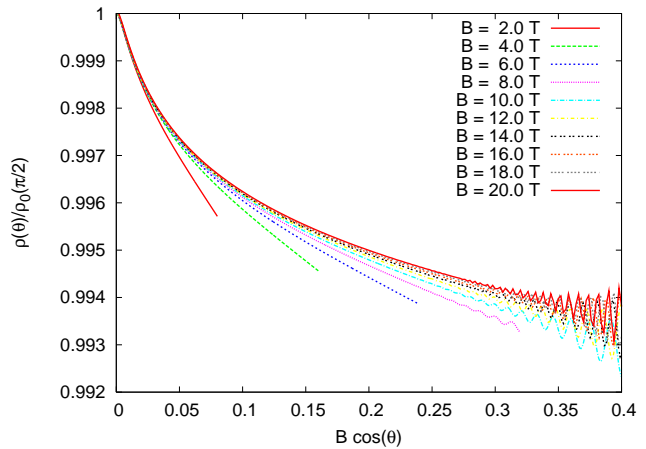}
\caption{Resistivity data from Fig.~\ref{fig:peak} plotted as a function of
the magnitude of the magnetic field parallel 
to the layers $B\cos\theta$.}
\label{fig:bperp}
\end{figure}

In Fig.~\ref{fig:bperp} we replot the data from Fig.~\ref{fig:peak} as
a function of $B_\perp = B\cos(\theta)$.  It is clear that as $B$ increases, the peak
becomes a function of $B_\perp$ only.  
We determined the width of the peak in 
two different ways: i) by choosing a 
fixed value of $\rho(\theta)/\rho(\theta=\frac{\pi}{2})$, 
and then determining the appropriate
$\Delta\theta$ as a function of $B$, and ii) by subtracting the (non-flat) background
and finding the value of $\Delta\theta$ corresponding to the half-maximum of the
peak.  In case i) $\Delta\theta \sim B^{-1}$ for essentially all values of $B$, whereas
for case ii) $\Delta\theta \sim B^{-1}$ for $B \gtrsim B_p \sim 6 \, {\rm T}$.

\begin{figure}[htb]
\insertpictwo{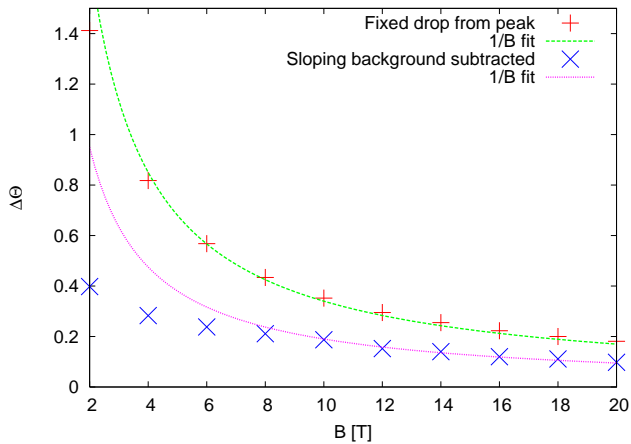}
\caption{Width of the peak in the resistivity about $\theta = 90^\circ$ as a function
of field for the data in Fig.~\ref{fig:peak}. The width is determined either by
a fixed drop in the resistance, or by subtracting the background.  Fits to both
procedures are shown.}
\label{fig:width}
\end{figure}

The results in Figs.~\ref{fig:bperp} and \ref{fig:width} are qualitatively in accord
with recent experiments on the resistivity in 
$\beta^{\prime\prime}$-(BEDT-TTF)$_4$[H$_3$OCr(C$_2$O$_4$)$_3$]CH$_2$Cl$_2$,
in which a peak that depends only on $B\cos(\theta)$ for small $\theta$ and a peak
width $\Delta\theta \sim 1/B$ were observed.\cite{coldea2} Additionally, features
in unpublished data\cite{kartsovnik}
on $\alpha$-[(ET)$_{1-x}$(BETS)$_x$]$_2$KHg(SCN)$_4$ with $x \simeq 0.03$,
 appear to be consistent with the picture presented here.

\subsection{Implications for experiment}
Inelastic scattering plays an important role
in determining when a resistivity peak will be observed. 
 A reasonable temperature 
dependence of the inelastic scattering rate 
implies that at high temperatures, the inelastic scattering time
is much less than the elastic scattering time $\tau_\phi \ll \tau_{11}$.
  At a temperature where the two become comparable, 
one might expect to see the 
first hints of a resistance peak which narrows and grows with decreasing temperature
until the peak width and height saturate.  We expect there will be a saturation when
$1/\tau_a$ reaches its low temperature limit $\sim \frac{1}{\tau_{11}} - \frac{1}{\tau_{12}}$
which is generally non-zero.  The visibility of the peak is largest when $\tau_a \gg 
\tau_{11}$, which will be true when $\tau_{12}$ is close in magnitude to $\tau_{11}$.

In concert with the magnetic field dependent signatures discussed above, the
temperature and magnetic field dependence of such a peak in the resistivity
should allow it to be distinguished from coherence peaks,
 whose width and height should have much weaker dependence on
 temperature and magnetic field.\cite{Moses}
As a consequence, these localization peaks also offer an opportunity to 
determine the inelastic scattering rate, independent of regular AMRO which
allow a determination of the elastic scattering rate, and quantum oscillations
which allow a determination of the total scattering rate through the
Dingle temperature.

Some questions that we have not addressed here but provide
interesting avenues for further exploration regard the effects of anisotropy
and the coherence of interlayer transport.  Our calculation was
for an isotropic Fermi surface with isotropic elastic scattering --
in general one can expect Fermi surface properties to vary on 
different parts of the Fermi surface, which would likely imply
that the nature of the peak for parallel fields will also 
depend on the orientation of the magnetic field within the plane.
This might allow the possibility of investigating the inelastic
scattering on different parts of the Fermi surface.  Secondly,
our derivation assumed weakly incoherent interlayer transport.
However, it appears likely to us that similar results should also hold for
coherent interlayer transport, and so the presence or absence of this
weak localization
peak is not necessarily evidence for or against coherent interlayer 
transport. The visibility of the peak will be stronger
in samples with relatively smaller values of $k_F l$, corresponding to
smaller values of $\tau$, but the criterion usually used to argue 
coherence of interlayer transport is comparing the hopping
amplitude, $t_\perp$, with the scattering rate $\frac{\hbar}{\tau}$.  

\section{Conclusions}
In conclusion, we have calculated the effect of the interlayer Cooperon
on AMRO in quasi-two dimensional layered metals.  The interlayer
Cooperon can give rise to a peak in the magnetoresistance for fields
parallel to the layers that behaves similarly to conventional weak localization.
The features of the peak depend sensitively on magnetic field and the
inelastic scattering time, which should allow it to be distinguished 
from coherence peaks seen in very low disorder layered metals.  This
potentially allows for the extraction of information about inelastic
scattering in layered metals, and gives another tool with which to 
learn about the physics of these systems.

\section{Acknowledgements}
The authors thank A. Coldea, 
R. Joynt, M. Kartsovnik, T. Kawamoto, T. Osada, B. Powell,
 and J. Singleton for helpful discussions.
We are particularly thankful to A. Coldea and M. Kartsovnik
for showing us their experimental data prior to publication.
M. Smith provided a critical reading of the manuscript.
This work was supported by an Australian
Research Council Discovery Project grant (R.H.M.), and 
by NSERC (M.P.K.). \\ \\

\begin{appendix}
\section{Calculation of $F(\bvec{B}_\parallel)$}
\label{sec:Fbparallel}
To calculate $F(\bvec{B}_\parallel)$ we need to 
perform an integral over four in-plane disorder averaged 
Green's functions
\begin{eqnarray}
F(\bvec{B}_\parallel) & = &
\int \frac{d^2\bvec{k}_1}{(2\pi)^2}
G^R(\bvec{k}_1) G^R\left(-\bvec{k}_1 \right)
\nonumber \\
& & \,\, \times
G^A\left(\bvec{k}_1 + \frac{ec}{2\hbar}\bvec{B}_\parallel\right)
G^A\left(-\bvec{k}_1 + \frac{ec}{2\hbar}\bvec{B}_\parallel \right) \nonumber \\
& = & \int \frac{d^2\bvec{k}_1}{(2\pi)^2}
\left(\frac{1}{E - \epsilon_k + i\eta}\right)^2
\frac{1}{E - \epsilon_{k-\beta} - i\eta}
\nonumber \\
& & \times
 \frac{1}{E - \epsilon_{k+\beta} - i\eta},
\label{eq:fbparallel_def}
\end{eqnarray}
with $\eta = \hbar/2\tau_{11}$, $\epsilon_k = \frac{\hbar^2 k^2}{2m^*}$, and
$\epsilon_{k-\beta} = \epsilon_k - \sqrt{\frac{2\epsilon_k}{m^*}}\beta \cos\phi +
\frac{\beta^2}{2m^*}$, with $\beta = \frac{1}{2}ecB_\parallel$.
Rewriting Eq.~(\ref{eq:fbparallel_def}) as an integral over energy and
performing the energy integral using contour integration, we get

\begin{eqnarray}
F(\bvec{B}_\parallel) & = &  \frac{m^*}{8\pi\hbar^2 \eta^3} \int_0^{2\pi}
 d\phi \left(1 - \frac{\beta^2}{4im^*\eta}\cos(2\phi)\right) \nonumber \\
& &  \times \frac{1}{\left(1 -
i\frac{\beta^2}{4\eta m^*} + i \frac{\beta\hbar k}{2\eta m^*}\cos\phi\right)^2} \\
& & \times \left.
\frac{1}{\left(1 - i\frac{\beta^2}{4\eta m^*} - i
\frac{\beta\hbar k}{2\eta m^*}\cos\phi\right)^2 }
\right|_{\epsilon_k = E + i\eta}.   \nonumber 
\end{eqnarray}
The second term in the denominators is much smaller than the third
term if $\beta \ll 2\hbar k_F$. This corresponds to
$B_\parallel \ll 4 \hbar k_F/(e c) \sim O(100 \, {\rm T})$, which is
always satisfied for physically realistic parameters.
We then have
\begin{equation}
{\rm Re} F(\bvec{B}_\parallel)
 = -\frac{m^*}{4\hbar^2 \eta^3}
f\left(\frac{\beta v_F}{\eta}\right).
\label{eq:fbparallel_two}
\end{equation}
where 
\begin{equation}
f(x) \equiv \frac{1}{2\pi}
\int_0^{2\pi} \frac{d\phi}
{\left(1+ \left(\frac{x}{2}\right)^2\cos^2 \phi \right)^2 }.
\end{equation}
Evaluating this by contour integration gives
Eqn. (\ref{eq:fx}).

\end{appendix}

%
%


\begin{thebibliography}{99}
\bibitem{aside}
As an aside, we note that
another interesting feature of the AMRO in these 
materials are that they are inverted with respect 
to the expectation for isotropic interlayer hopping.
Such a situation can occur when the interlayer hopping
$t_\perp(\phi)$ varies significantly around the Fermi surface 
in the appropriate way.\cite{yagi}  It is interesting
to note that the the planes of ET molecules in adjacent layers are
twisted with respect to each other by 62 degrees,\cite{coldea}
which may be responsible for such a intralayer variation
of the interlayer hopping integral.
\end{thebibliography}
\end{document}